\documentclass[preprint,showpacs,preprintnumbers,amsmath,amssymb,nofootinbib]{revtex4}

\usepackage{amsthm,amscd,amsbsy,array}
\usepackage{bm}

\usepackage{graphics,graphicx,xcolor}

\newcommand{\gb}{\colorbox{green}}

\newenvironment{redtext}{\color{red}}{\ignorespacesafterend}
\newenvironment{bluetext}{\color{blue}}{\ignorespacesafterend}

\newcommand{\bblue}{\begin{bluetext}}
\newcommand{\eblue}{\end{bluetext}}
\newcommand{\bred}{\begin{redtext}}
\newcommand{\ered}{\end{redtext}}

\numberwithin{equation}{section}

\let\ssection=\section
\renewcommand{\section}{\setcounter{equation}{0}\ssection}

\newcommand{\cA}{{\mathcal{A}}}

\newcommand{\bb}{{\bf b}}

\newcommand{\bone}{\boldsymbol{1}}

\newcommand{\bc}{{\mathbf{c}}}

\newcommand{\diag}{\mathrm{diag}}

\newcommand{\dgamma}{\dot{\gamma}}

\newcommand{\ddchi}{\ddot{\chi}}

\newcommand{\rg}{\mathrm{g}}

\newcommand{\bk}{\mathbf{k}}

\newcommand{\bp}{{\bf p}}

\newcommand{\bx}{{\bm{x}}}

\newcommand{\Tr}{\mathrm{Tr}}

\newcommand{\bX}{{\bf X}}

\def\smallover#1/#2{\hbox{$\textstyle\frac{#1}{#2}$}} %
\def\Rarrow{\quad\Rightarrow\quad}

\def\bp{{\bm{p}}}

\def\benu{\begin{enumerate}}
\def\eenu{\end{enumerate}}
\def\beq{\begin{equation}}
\def\eeq{\end{equation}}
\def\beqa{\begin{eqnarray}}
\def\eeqa{\end{eqnarray}}

\def\barray{\left(\begin{array}}
\def\earray{\end{array}\right)}
\def\barraynb{\begin{array}}
\def\earraynb{\end{array}}

\def\?{\quad{\gb{\fbox{\texttt{?}}\;}}\quad}

\def\v0{\mathbf{0}}

\def\beq{\begin{equation}}
\def\eeq{\end{equation}}
\def\bea{\begin{eqnarray}}
\def\eea{\end{eqnarray}}

\def\6{\partial}
\def\7{\tilde}
\def\8{\widehat}
 \def\bx{{\bf x}}

\newcommand{\const}{\mathop{\rm const.}\nolimits}
\newcommand{\half }{\frac{1}{2}}

\def\smallover#1/#2{\hbox{$\textstyle\frac{#1}{#2}$}} %
\def\smallcirc{{\raise 0.5pt \hbox{$\scriptstyle\circ$}}}
\def\2{{\smallover1/2}}

\let\ssection=\section
\renewcommand{\section}{\setcounter{equation}{0}\ssection}

\begin{document} 

\preprint{[arXiv:1704.05997v4 [gr-qc]].
}

\title{The Memory Effect for Plane Gravitational Waves
\\[6pt]}

\author{
P.-M. Zhang$^{1}$\footnote{e-mail:zhpm@impcas.ac.cn},
C. Duval$^{2}$\footnote{
mailto:duval@cpt.univ-mrs.fr},
G. W. Gibbons$^{3,4,5}$\footnote{
mailto:G.W.Gibbons@damtp.cam.ac.uk},
P. A. Horvathy$^{1,4}$\footnote{mailto:horvathy@lmpt.univ-tours.fr},
}

\affiliation{
$^1$Institute of Modern Physics, Chinese Academy of Sciences, Lanzhou, China
\\
$^2$ Aix Marseille Univ, Universit\'e de Toulon, CNRS, CPT, Marseille, France
\\
$^3$D.A.M.T.P., Cambridge University, U.K.
\\
$^4$Laboratoire de Math\'ematiques et de Physique
Th\'eorique,
Universit\'e de Tours,
France
\\
$^5$LE STUDIUM, Loire Valley Institute for Advanced Studies, Tours and Orleans France
}

\date{\today}


\begin{abstract}
We give an account of the gravitational memory effect
in the presence of the exact plane wave solution of Einstein's vacuum equations. This allows an elementary but exact description of the soft gravitons and how their presence may be detected by observing the motion of freely falling particles. The theorem of Bondi and Pirani on caustics (for which we present a new proof) implies that the asymptotic relative velocity is constant but not zero, in contradiction with the permanent displacement claimed by Zel'dovich and Polnarev. A non-vanishing asymptotic relative velocity might be used to detect gravitational waves through the ``velocity memory effect", considered by Braginsky, Thorne, Grishchuk, and Polnarev.
\\[8pt]
  Phys.\ Lett.\ B {\bf 772} (2017) 743.
  doi:10.1016/j.physletb.2017.07.050
\end{abstract}

\maketitle


\section{Introduction}\label{Intro}

By  ``gravitational memory effect''  is meant that  a system of freely falling particles
(viewed as detectors) initially at relative rest are displaced
after the passing of a burst of gravitational radiation  \cite{ZelPol,BraGri}.
 The  memory effect has risen to prominence recently because of the possibility
of  its observation using  LISA \cite{Fava,Lasenby} and/or aLIGO \cite{Lasky}, and is also  relevant to recent work by Hawking, Perry, and Strominger \cite{PeHaStPRL} on soft gravitons and black holes.

The original proposal of Zel'dovich and Polnarev \cite{ZelPol} claimed that
 detectors originally at rest will be simply displaced and
will have vanishingly small relative velocity, in
 contrast with  later findings of Bondi and Pirani \cite{BoPi89}. A ``velocity memory effect" highlighted by non-vanishing asymptotic relative velocity was
 considered by   Braginsky, Grishchuk, Thorne, and Polnarev
 \cite{BraGri, BraTho, GriPol} and more recently, by Lasenby \cite{Lasenby}.

The analysis of Braginsky and Grishchuk \cite{BraGri}, who also coined the expression ``memory effect'', is concerned with the motion of test masses  moving in a weak gravitational wave at the linear level.
Non-linear generalizations were considered in
 \cite{BlaDam,Christo,Thor,Harte12}.
In the neighborhood of a detector, far from the source, we may approximate the  wave  by an exact plane wave.
 Carroll symmetry  \cite{Sou73,Carroll4GW} provides us with a description of the trajectories, leading  to a new proof of the Bondi-Pirani theorem \cite{BoPi89}. Our results might have some relevance for detecting gravitational waves through the ``velocity memory effect"  \cite{BraGri, BraTho, GriPol,LongMemory,Lasenby}.
\goodbreak

\section{Plane Gravitational Waves}\label{Planewaves}


Plane gravitational waves are often described in
\emph{Brinkmann Coordinates} (B) \cite{Bri,BoPiRo58} by the metric
\beq
\rg =\delta_{ij}\,dX^idX^j+2dUdV+K_{ij}(U){X^i}{X^j}\,dU^2,
\label{Bplanewave}
\eeq
where the symmetric and traceless $2\times2$ matrix
$K(U)=\left(K_{ij}(U)\right)$ characterizes the profile of the wave.
 In this Letter we only consider linearly polarized waves with
\beq
K_{ij}(U){X^i}{X^j}=
\half\cA(U)\Big((X^1)^2-(X^2)^2\Big),
\label{genBrink}
\eeq
where $\cA(U)$ is an arbitrary function. Our particular interest lies in ``sandwich waves" i.e. such that $K(U)\neq0$ only for $U_i\leq U\leq U_f$ \footnote{Our conventions are the opposite of that of Bondi and Pirani \cite{BoPi89}.}.

The metric (\ref{Bplanewave}) is an example of
  a ``Bargmann'' space which provides a general framework for describing non-relativistic dynamical systems~: any natural non-relativistic  system with an $n$ dimensional  configuration space may be ``Eisenhart-lifted'' to a system of null geodesics in an $(n+2)$-dimensional Lorentzian spacetime endowed with a covariantly constant null Killing vector field $\partial_V$ \cite{Eisenhart,DBKP}.
From the Bargmann point of view, (\ref{Bplanewave})-(\ref{genBrink}) describe a non-relativistic particle subjected to an (attractive or repulsive) harmonic  anisotropic oscillator potential, (\ref{genBrink}).

\begin{figure}[ht]
\hskip-7mm
\includegraphics[width=0.4\textwidth]{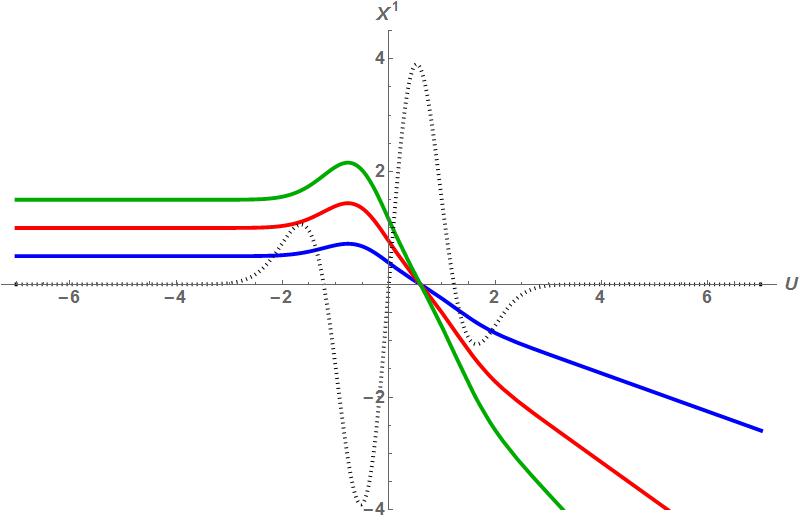}
\includegraphics[width=0.4\textwidth]{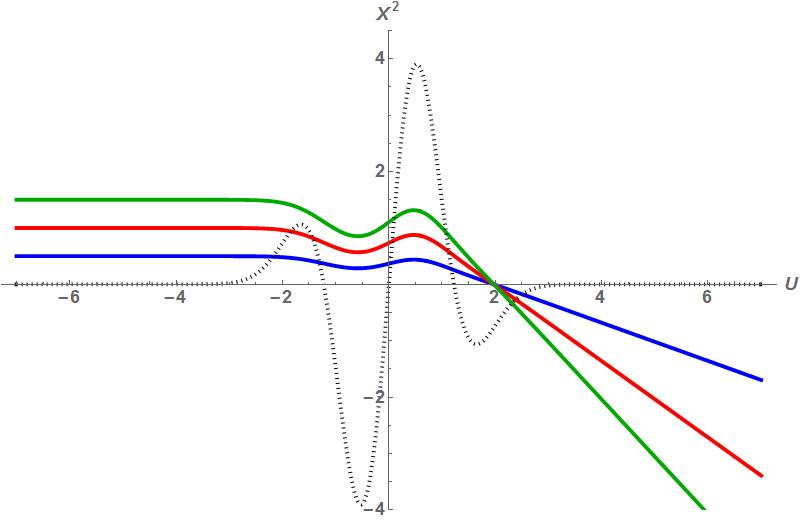}
\\
\vskip-3mm
\caption{\textit{Geodesics in Brinkmann coordinates for particles initially at rest  for profile $\cA(U)
$\, in (\ref{Kd3}), modelling gravitational collapse.
The trajectories are focused where the effective potential is attractive, and are pushed apart where it is repulsive. In the flat afterzone, the acceleration vanishes.}}
\label{d3traj}
\end{figure}

Returning to gravitational radiation and following the suggestion of \cite{Gibbons:1972}, the wave produced by gravitational collapse  might be modelled by the (approximate) sandwich wave,
\beq
\cA(U)
 =\frac{1}{2}\,\frac{d^3(e^{-U^2})}{dU^3}\,
\label{Kd3}
\eeq
shown in fig.\ref{d3traj}.

In a (naive) model the ``detector'' could consist of freely falling  particles initially at relative rest.
This  applies to the mirrors of the LIGO interferometers
of those  attached to the three satellites making up the proposed LISA detector
\cite{Lasenby}.

The profile function in the geodesic equations\footnote{$V$ satisfies a more complicated equation, (IV.2.c) in \cite{LongMemory}.},
\beq
\dfrac {d^2\! \bX}{dU^2} - \half\diag(\cA,-\cA)\,\bX
= 0
\label{Bgeoeq}
\eeq
acts as an effective  potential which is attractive or repulsive depending on the sign of $\cA$. Eqn. (\ref{Bgeoeq}) can be solved numerically, yielding Fig.\ref{d3traj}. \vskip -2mm

\begin{figure}[h]\hskip-6mm
\includegraphics[width=0.4\textwidth]{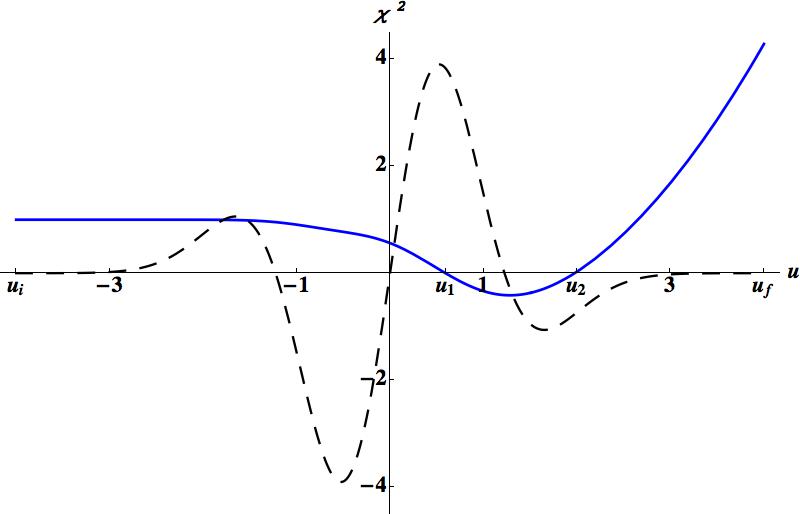}
\includegraphics[width=0.4\textwidth]{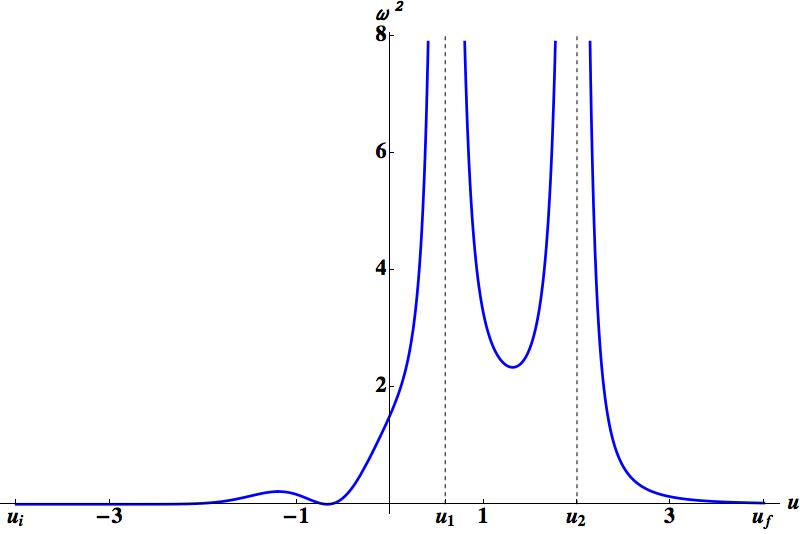}
\caption{\textit{The numerical solution of eqn. (\ref{chiSL})  for the ``collapse profile'' (\ref{Kd3}) indicates that the coordinates become singular  when  $u_1 = 0.593342$
and when $u_2 = 1.97472$.}}
\label{chifig}
\end{figure}

Gravitational waves can also be discussed in \emph{Baldwin-Jeffery-Rosen Coordinates} (BJR) \cite{BaJe,Ro}, for which
\beq
\rg=a_{ij}(u)\,dx^idx^j+2du\,dv
\label{BJRmetrics}
\eeq
with $a(u)=(a_{ij}(u))$ a positive matrix. The BJR coordinates are not harmonic and typically not global,  exhibiting coordinate singularities \cite{Ro,BoPiRo58,Sou73}.
Let us recall the proof \cite{Sou73,LongMemory}.
 Putting $\chi=\big(\det{a}\big)^{\frac{1}{4}}>0$ and
$\omega^2(u)=\frac{1}{8}\Tr\left((\gamma^{-1}\dgamma)^2\right)$ where $\gamma=\chi^{-2}a$, the vacuum Einstein equations reduce to the Sturm-Liouville equation
\beq
\ddchi(u)+\omega^2(u)\chi(u)=0,
\label{chiSL}
\eeq
 see eqn (II.22) of \cite{LongMemory}. Then (\ref{chiSL}) implies
 that  $\ddot{\chi}<0$, which
in turn implies the vanishing of $\chi$ for some $u_1 > u_i$,
\beq
\chi(u_1)=0,
\label{chizero}
\eeq
 signalling a singularity of the metric (\ref{BJRmetrics})
and illustrated by fig.\ref{chifig}.

\goodbreak
Plane gravitational waves  have a 5-parameter isometry group \cite{BoPiRo58,Sou73,BoPi89} identified
 as the \emph{Carroll group in $2+1$ dimensions with broken rotations} \cite{Carroll4GW} and implemented in BJR coordinates as follows~: $u$ is fixed, and
\begin{subequations}
\begin{align}
\label{xCarrol}
\bx&\to\bx+H(u)\,\bb+\bc,
\\
v&\to v-\bb\cdot\bx - \2\bb\cdot{}H(u)\,\bb+f,
\label{vCarrol}
\end{align}
\label{genCarr}
\end{subequations}
where $H$ is the symmetric $2\times 2$ matrix
 calculated using the transverse-space matrix,
 \beq
H(u)=\int^u_{u_0}\!\!a(t)^{-1} dt.
\label{Hmatrix}
\eeq
Then Noether's theorem provides us with 5 conserved quantities associated with these isometries  \cite{Sou73,Carroll4GW},
\beq
\bp\!=\! a(u)\,\dot{\bx},
\quad
\bk\!=\!\bx(u)-H(u)\,\bp,
\label{CarCons}
\eeq
 interpreted as  conserved \emph{linear and  boost-momentum}, supplemented by $m=\dot{u}=1$.
An extra constant of the motion is
$
e=\half\rg_{\mu\nu}\,\dot{x}^\mu\dot{x}^\nu.
$
Geodesics are timelike, lightlike, or spacelike if  $e$ is
negative, zero, or positive, respectively. The geodesics are thus given by \cite{Sou73,Carroll4GW}\vskip-10mm
\begin{subequations}
\begin{align}
\label{xtraj}
\bx(u)&=H(u)\,\bp+\bk,
\\
\label{traj}
v(u)&=-\half \bp\cdot H(u)\,\bp + e\,u+ d,\quad
\end{align}
\label{BJRGeo}
\end{subequations}
 with $d$ a constant. The trajectories can be determined once the matrix-valued function $H(u)$ in (\ref{Hmatrix}) had been calculated.

The relation between the two coordinates systems is given by \cite{Gibb75,Carroll4GW}
\beq
{\bX} =P(u)\,\bx,
\;\;\;
V=v-\frac{1}{4}\bx\cdot\dot{a}(u)\bx\,,
\label{BBJRtrans}
\eeq
where $U=u$  \footnote{The notation $u$ or $U$ is used only to distinguish which coordinates we are working with.},  $a(u)=P(u)^{T}{}P(u)$ and
\beq
\ddot{P}=K\,P \,,
\qquad
P^{T}\dot{P}-\dot{P^{T}}P=0 \,.
\label{SLB+2}
\eeq

Now we consider a sandwich wave.
In the ``beforezone'' $U<U_i$ $K=0$ and therefore
eqn. (\ref{SLB+2}) can be solved by $P(u)=\bone$ so that the Brinkmann and BJR coordinates coincide.

The crucial fact for us is that, by (\ref{CarCons}),   the  momentum of a particle at rest  vanishes, $\bp=0$ for $u=U\leq U_i$ --- and thus for \emph{all} $u=U$ including the ``inside zone'' $U_i \leq U \leq U_f$ and the ``afterzone'' $U_f \leq U$, because $\bp$ is conserved. By (\ref{xtraj}) such geodesics are, \emph{for any $H$ i.e. for any transverse metric $a$} and therefore \emph{for any profile $\cA$}, simply
\begin{equation}
\bx(u)=\bx_0,
\quad
v(u)=e\,u+v_0.
\label{SpecGeod}
\end{equation}
Thus the  \emph{particles initially at rest are, when described in BJR coordinates, undisturbed by the passage of the wave~\footnote{
If the initial momentum does not vanish, though, $\bp\neq0$, then the trajectories are obviously not simple straight lines; they can be determined by calculating $H(u)$ in (\ref{Hmatrix}), cf. \cite{LongMemory}. }}~.

In Brinkmann coordinates both the waves and the geodesics are global with no singularity whatsoever. Solving the equations (\ref{SLB+2})  numerically for $P$ yields,
\beq
\bX(U)=P(u)\bx^0,
 \qquad \bx^0=\const
\label{XPx}
\eeq
cf. (\ref{BBJRtrans}). Those complicated-looking trajectories in B coordinates in Fig.\ref{d3traj}
are recovered: the plots overlap perfectly \emph{up to the point where the BJR coordinate system becomes singular}; all the curling comes from $P(u)$.

Our plots show that, consistently with the theorem of Bondi and Pirani \cite{BoPi89}, \emph{after the wave has passed the worldlines will first be focused in a finite and initial-position-independent  time,
 and then diverge with \emph{constant but non-zero asymptotic velocity}} --  contradicting the claim of Zel'dovich and Polnarev \cite{ZelPol}.

We can actually prove the  Bondi-Pirani theorem independently using our framework.
First we note that  $\cA \approx 0$  for large values of $U$; then  it follows from the geodesic equations (\ref{Bgeoeq})  that the acceleration vanishes,  implying constant asymptotic velocity \footnote{Another proof using BJR coordinates is presented in \cite{LongMemory}. }.

 The geodesics we are interested in are of the form (\ref{XPx});
in the beforezone $U\leq U_i$ we have $P=\bone$ and the geodesics are at rest, $\bX(U)=\bx^0$.
Let us now consider two such special geodesics, $\bX_a(U)$ and $\bX_b(U)$; we show that there exist $\bx_a^0\neq \bx_b^0$ such that the two geodesics meet,
$
\bX_a(U_1)=\bX_b(U_1)
$
for some finite and  universal value $u_1> u_i$.
$\bX_{a,b}(U) = P(u) \bx_{a,b}^0$, so we need
 $\bx^0=\bx_a^0-\bx_b^0 \neq 0$ such that
$P(u_1) \bx^0 = 0$, which
happens when $P(u_1)$ is a singular matrix,
\beq
\det (P(u_1)) = 0.
\label{detP0}
\eeq
But
$
a= P^TP$ so that $\det a = (\det P)^2 = \chi^4
$
and therefore $\det P = 0$ iff $\chi=0$. However, as recalled above, (\ref{chiSL}) implies  that $\chi$ necessarily vanishes at some $u_1>u_i$ -- at the same value where the BJR coordinates become singular. This is not an accident: the singular BJR coordinate system can only be regularized by a singular transformation.
 Thus $P(u_1)$ has a kernel and all geodesics for which $\bx_a^0-\bx_b^0$ belongs to the kernel will meet at the same $u_1$.
If $\bx_a^0 - \bx_b^0$ does not belong to the kernel, then we get a caustic. For a linearly polarized wave as in
(\ref{Bplanewave})-(\ref{genBrink}),
$P=\diag(P_{11},P_{22})$ and the $x^1$ resp $x^2$ coordinates meet at critical (caustic) point when either $P_{11}$ or P$_{22}$ vanishes, which is where BJR coordinates become singular, cf. fig. \ref{d3traj}.

\begin{figure}[h] \hskip -6mm
\includegraphics[width=0.6\textwidth]{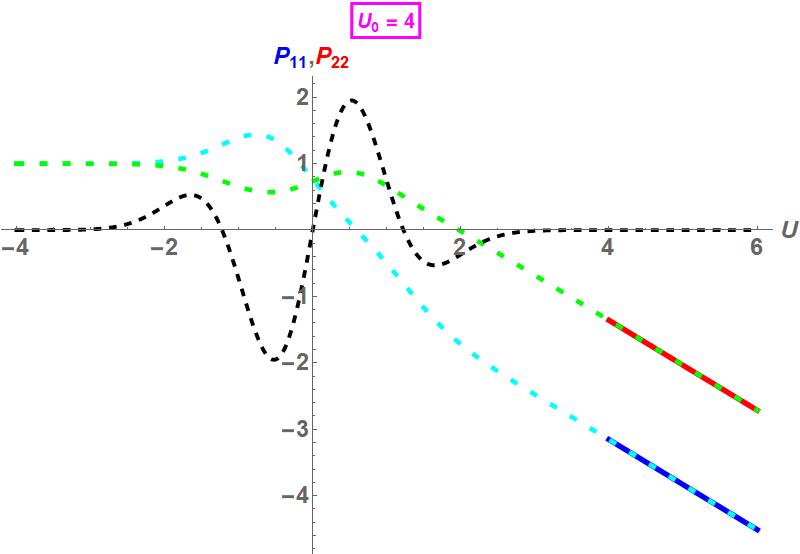}
\vskip-5mm
\caption{\textit{The analytic (heavy line) and numerical  (dashed line) solutions overlap perfectly in the afterzone $u\geq u_0=4$.
}}
\label{AZu4}
\end{figure}

In the flat afterzone $u \geq u_f$, in particular, we have the exact analytic solution \cite{Sou73,LongMemory}
\beq
P(u)= \big(\bone + (u-u_0)c_0\big)a_0^{1/2} \Rarrow
a(u)= a_0^\half \left[\bone+(u-u_0)c_0\right]^2 a_0^\half,
\label{Paafter}
\eeq
where we have chosen $u_0=u_f,\,a_0=a(u_0), \,
c_0=\half a_0^{-1/2}\dot{a}(u_0)a_0^{-1/2}\neq0$
 \cite{Sou73, LongMemory}. Setting $\chi_0 =\chi(u_0)$,
\beq
 \chi(u) = \chi_0 \left[1+(u-u_0)\Tr(c_0)+(u-u_0)^2 \det(c_0)\right]^\half.
\eeq
Then the requirement of continuity implies
$\Tr(c_0) = 2\,{\dot\chi_0}/{\chi_0},$
valid as long as $\chi\neq0$.
Thus $\Tr(c_0)\neq0$ unless $\dot\chi_0=0$.
Combining (\ref{XPx}) and  (\ref{Paafter}) yields  the trajectory in the afterzone, plotted in fig.\ref{AZu4}.
The asymptotic velocity is  inferred from (\ref{Paafter}),
\beq
\dot{\bX}(u)=c_0a_0^{1/2}\bx^0=\half a_0^{-1/2}\dot{a}(u_0)\bx^0=\const
\label{aftervel}
\eeq
Thus $\dot{\bX}_a\neq \dot{\bX}_b$ for $\bx_a^0\neq \bx_b^0$ in general, as confirmed by numerical calculations. At last, the linear form of $P$ in (\ref{Paafter}) implies that each component of $P$  can have at most one caustic in the afterzone.

\section{Illustration by Tissot diagram}

\begin{figure}[h]
\includegraphics[width=0.6\textwidth]{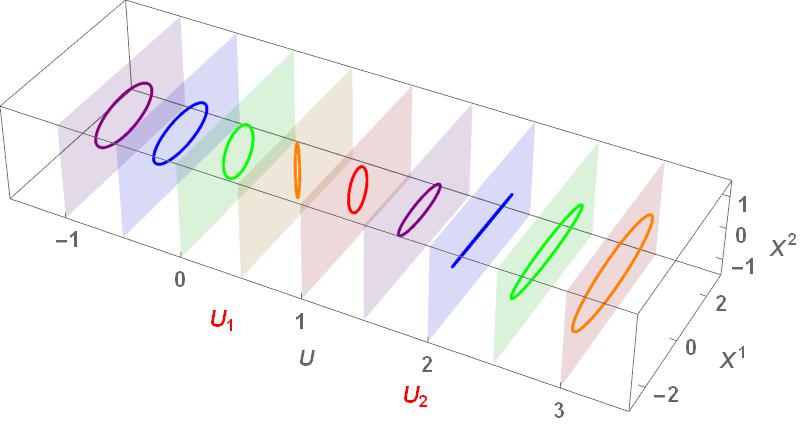}
\\
 \vskip -4mm
\caption{\textit{Tissot diagram for the ``collapse profile'' (\ref{Kd3}). The initial circle is deformed to an ellipse which at $u=u_1$ degenerates to a line segment, and so on.
}} \vskip -3mm
\label{Tissotfig}
\end{figure}

The action of gravitational waves  on a ring of freely falling particles can be illustrated by a series of time-frames showing how the ring is squashed and stretched as the wave passes over it \cite{MiThoWh}. An initially (i.e. before the pulse) circular
disc of geodesics,
$\bX\cdot{}\bX \le 1$
for $U < U_i$, projects to a time-independent
circle, $\bx\cdot{}\bx \le 1 $,
which extends to all $u$, i.e., during and after the sandwich, $u\geq u_i$. However
their inverse images in Brinkmann coordinates are time dependent ellipses,
\beq
1=\bx\cdot\bx=
 \bX\cdot{} (P P^{T})^{-1} \bX
\eeq
as shown in Fig. \ref{Tissotfig}.

\section{Soft gravitons}

We end with a description  of the flat spacetime
 after the burst as being dressed by soft gravitons, a concept which
  plays a central role in the considerations of \cite{PeHaStPRL}.
 Before the burst  $a_{ij}=\delta_{ij}$, and the BJR coordinates coincide
 with standard inertial coordinates in Minkowski spacetime.
 After the burst $a_{ij}$ will be $u$-dependent but nevertheless will
 have vanishing curvature. Thus there exist
 coordinates $\hat u,  \hat v, \hat {\bx}$  such that the $2$-metric is
 $\hat{a}_{ij} =\delta_{ij}$. The transformation relating  $ u, v, {\bx}$ to the $\hat{u},  \hat{v}, \hat{\bx}$
 coordinates is given by (\ref{BBJRtrans}) with
 $\ddot P=0$, as  follows from the vanishing of $K$.
 In the  quantum field theory language adopted by Hawking et al. in \cite{PeHaStPRL}, the coordinate transformation
 is  a  ``gauge  transformation  which
 does not tend to the identity at infinity''. In quantum field theory
 it is customary to distinguish states which so differ and regard them as
 containing  ``soft'' [i.e. zero-energy] quanta. In the present case the quanta are gravitons given by pp-waves,  whose  metric is given by (\ref{BJRmetrics}).

\section{Conclusion}

We found  that test particles initially at rest move, after the wave has passed, with \emph{constant but non-zero relative velocity}, contradicting  Zel'dovich and  Polnarev \cite{ZelPol}
but consistently with Bondi and Pirani \cite{BoPi89}, whose theorem we proved independently.  The non-vanishing asymptotic relative velocity might allow to observe the \emph{velocity memory effect} by  improved Doppler tracking
\cite{BraGri,BraTho,GriPol} \footnote{P. Lasky kindly informed us that aLIGO may be able to observe  non-permanent displacements in a not-too-distant future.}.

The intuitive explanation is that (\ref{Bplanewave})-(\ref{genBrink})  is (as said above), the Bargmann metric of an anisotropic oscillator with time-dependent frequencies, one of which is attractive and the other repulsive. The components in the repulsive sector are pushed apart whereas in the attractive sector they are focused  and then meet where the matrix $P$ is singular.

\goodbreak
\begin{acknowledgments}
GWG would like to thank the
{\it Laboratoire de Math\'ematiques et de Physique Th\'eorique de l'Universit\'e de Tours}  for hospitality and the  {\it R\'egion Centre} for a \emph{``Le Studium''} research professor\-ship.
PH thanks  the \emph{Institute of Modern Physics} of the Chinese Academy of Sciences in Lanzhou  for hospitality. Support by the National Natural Science Foundation of China (Grant No. 11575254) is acknowledged.
The authors are grateful to Shahar
 Hadar, Abraham Harte,  Anthony Lasenby, Paul Lasky and Malcolm Perry for  helpful exchanges.
\end{acknowledgments}
\goodbreak


\end{document}